\documentclass[preprint]{aastex}

\shorttitle{29 pulsars discovered in the PALFA Survey}
\shortauthors{A. G. Lyne et al.}

\begin{document}

\title{Timing of 29 Pulsars Discovered in the PALFA Survey}

\author{
A.~G.~Lyne\altaffilmark{1}, 
B.~W.~Stappers\altaffilmark{1},
S.~Bogdanov\altaffilmark{2},
R.~D.~Ferdman\altaffilmark{3},
P.~C.~C.~Freire\altaffilmark{4},
V.~M.~Kaspi\altaffilmark{3},
B.~Knispel\altaffilmark{5,6},  
R.~Lynch\altaffilmark{3}, 
B.~Allen\altaffilmark{5,6,7},
A.~Brazier\altaffilmark{8},
F.~Camilo\altaffilmark{9},
F.~Cardoso\altaffilmark{6},
S.~Chatterjee\altaffilmark{8},
J.~M.~Cordes\altaffilmark{8},
F.~Crawford\altaffilmark{10},
J.~S.~Deneva\altaffilmark{11},
J.~W.~T.~Hessels\altaffilmark{12,13},
F.~A.~Jenet\altaffilmark{14}, 
P.~Lazarus\altaffilmark{4}, 
J.~van~Leeuwen\altaffilmark{12,13}, 
D.~R.~Lorimer\altaffilmark{15},
E.~Madsen\altaffilmark{3},
J.~McKee\altaffilmark{1},
M.~A.~McLaughlin\altaffilmark{15},
E.~Parent\altaffilmark{3},
C.~Patel\altaffilmark{3},
S.~M.~Ransom\altaffilmark{16},
P.~Scholz\altaffilmark{3},
A.~Seymour\altaffilmark{4},
X.~Siemens\altaffilmark{6}, 
L.~G.~Spitler\altaffilmark{4},
I.~H.~Stairs\altaffilmark{17,3}, 
K.~Stovall\altaffilmark{18,19}, 
J.~Swiggum\altaffilmark{6},
R.~S.~Wharton\altaffilmark{8},
W.~W.~Zhu\altaffilmark{4},
C.~Aulbert\altaffilmark{5,7},
O.~Bock\altaffilmark{5,7},
H.-B.~Eggenstein\altaffilmark{5,7},
H.~Fehrmann\altaffilmark{5,7},
B.~Machenschalk\altaffilmark{5,7}
}

\altaffiltext{1}{Jodrell Bank Centre for Astrophys., School of Phys. and Astr., Univ. of Manchester, Manch., M13 9PL, UK} 
\altaffiltext{2}{Columbia Astrophysics Laboratory, Columbia Univ., New York, NY 10027, USA} 
\altaffiltext{3}{Dept.~of Physics and McGill Space Institute, McGill Univ., Montreal, QC H3A 2T8, Canada}
\altaffiltext{4}{Max-Planck-Institut f\"ur Radioastronomie, Auf dem H\"ugel 69, D-53121 Bonn, Germany} 
\altaffiltext{5}{Max-Planck-Institut f\"ur Gravitationsphysik, D-30167 Hannover, Germany}
\altaffiltext{6}{Physics Dept., Univ. of Wisconsin - Milwaukee, 3135 N. Maryland Ave., Milwaukee, WI 53211, USA}
\altaffiltext{7}{Leibniz Universit{\"a}t Hannover, D-30167 Hannover, Germany}
\altaffiltext{8}{Dept. of Astronomy, Cornell Univ., Ithaca, NY 14853, USA} 
\altaffiltext{9}{SKA South Africa, Pinelands, 7405, South Africa} 
\altaffiltext{10}{Dept. of Physics and Astronomy, Franklin and Marshall College, Lancaster, PA 17604-3003, USA} 
\altaffiltext{11}{National Research Council, resident at the Naval Research Laboratory, Washington, DC 20375, USA}
\altaffiltext{12}{ASTRON, The Netherlands Institute for Radio Astronomy, Postbus 2, 7990 AA, Dwingeloo, The Netherlands} 
\altaffiltext{13}{Anton Pannekoek Institute for Astronomy, Univ. of Amsterdam, Science Park 904, 1098 XH Amsterdam, The Netherlands}
\altaffiltext{14}{Center for Gravitational Wave Astronomy, Univ. Texas - Brownsville, TX 78520, USA} 
\altaffiltext{15}{Dept. of Physics, West Virginia Univ., Morgantown, WV 26506, USA} 
\altaffiltext{16}{NRAO, Charlottesville, VA 22903, USA} 
\altaffiltext{17}{Dept.~of Physics and Astronomy, Univ.~of British Columbia, Vancouver, BC V6T 1Z1, Canada} 
\altaffiltext{18}{NRAO, PO Box 0, Socorro, NM 87801, USA}
\altaffiltext{19}{Dept. of Physics and Astronomy, Univ.~of New Mexico, NM 87131, USA}

\begin{abstract}

We report on the discovery and timing observations of 29 distant
long-period pulsars found in the ongoing Arecibo L-band Feed
Array pulsar 
survey.  Following discovery with the Arecibo Telescope, confirmation
and timing observations of these pulsars over several years at Jodrell
Bank Observatory have yielded high-precision positions and
measurements of rotation and radiation properties. We have used
multi-frequency data to measure the interstellar scattering properties
of some of these pulsars.  Most of the pulsars have properties that
mirror those of the previously known pulsar population, although four
show some notable characteristics.  PSRs J1907+0631 and J1925+1720 are
young and are associated with supernova remnants or plerionic nebulae:
J1907+0631 lies close to the center of SNR G40.5$-$0.5, while
J1925+1720 is coincident with a high-energy ${\it Fermi}$ $\gamma$-ray
source. One pulsar, J1932+1500, is in a surprisingly eccentric,
199-day binary orbit with a companion having a minimum mass of 0.33
M$_\odot$.  Several of the sources exhibit timing noise, and two, PSRs
J0611+1436 and J1907+0631, have both suffered large glitches, but with
very different post-glitch rotation properties. In particular, the
rotational period of PSR~J0611+1436 will not recover to its pre-glitch
value for about 12 years, a far greater recovery timescale than seen
following any other large glitches.

\end{abstract}

\section*{}\eject  

\section{Introduction}\label{sec:intro}

The Pulsar Arecibo L-Band Feed Array (PALFA) project is a deep pulsar
survey of low Galactic latitudes being conducted at the 305-m William
E.\ Gordon Telescope at the Arecibo Observatory
\citep{cfl+06,lab+12,nab+13,lbh+15}. The survey's relatively high
operating frequency of 1420~MHz makes it particularly sensitive to
pulsars over a far larger volume of the Galaxy than previous surveys,
notably to those pulsars with high dispersion measures (DMs) and short
spin periods \citep{csl+12}.

The primary aims of the survey include the discovery of binary pulsars
which might provide tests of strong-field gravitation theories and the
determination of the neutron star equation of state
\citep{sta03,kbc+04}, the discovery of millisecond pulsars which might
be included in pulsar timing arrays in the quest for gravitational
waves \citep{haa+10}, and to better determine the Galactic
distribution of the pulsar population \citep{lbdh93,acc02} and the
Galactic interstellar medium \citep{cl02,han04}. The properties of
PALFA also make it an ideal instrument for discovering young pulsars
in the Galactic plane that may have associations with supernova
remnants and other high-energy phenomena and are also likely to have
notable timing properties for studying neutron star interiors.

In a previous paper, \citet{nab+13} presented timing measurements of a
first tranche of 35 relatively slowly rotating pulsars discovered in
the PALFA survey prior to 2008. Here, we describe timing observations
of a further 29 such pulsars discovered up to 2013.  These
observations were made primarily with the Lovell Telescope at Jodrell
Bank Observatory, with a few observations using the 100-m Robert
C. Byrd Green Bank Telescope.  Following a brief description of the
survey in Section~\ref{sec:survey} and presentation of the observations in
Section~\ref{sec:obs}, we give the results of timing analyses of the pulsars
in Section~\ref{sec:timing} and provide their flux densities and pulse
profiles in Section~\ref{sec:flux}.  We also discuss a small number of the
pulse profiles which show clear evidence for interstellar multipath
scattering.  In Sections~\ref{sec:young}--\ref{sec:glitches}, we discuss
several notable pulsars: two are young, one is in a binary system, and
two display large glitches in their rotation rate.  We present our
conclusions in Section~\ref{sec:conclusions}.

\section{The PALFA survey}\label{sec:survey}

Details of the PALFA survey are described in the following papers: 
\citet{cfl+06}, \citet{vcl+06}, \citet{lab+12}, \citet{nab+13},
\citet{slm+14} and \citet{lbh+15}.  We therefore give only a synopsis
here. 

The survey includes the regions of the Galactic plane
($|b|<5^{\circ}$; $32^{\circ} \lesssim \ell \lesssim 77 ^{\circ}$ and
$168{^\circ} \lesssim \ell \lesssim 214^{\circ}$ ) that are accessible
with the Arecibo Telescope.  Each pointing of the 7-beam ALFA receiver
had a dwell time of 268\,s and, for observations prior to 2009, a
bandwidth of 100 MHz was channelized and digitized using the Wideband
Arecibo Pulsar Processor (WAPP) 3-level autocorrelator system
\citep{dsh00}.  The dual polarizations were summed to give 256
total-intensity channels sampled every 64~$\mu$s. Subsequent
observations used the wider bandwidth of 322 MHz which was channelized
to 960 channels which were sampled every 65.5 $\mu$s using the Mock
spectrometers \citep{lbh+15}. Of the 29 pulsars reported in this
paper, 3 were discovered using the WAPP spectrometers and 26 using the
Mock spectrometers.

The large volume of data that is acquired has meant that several
analysis pipelines were used to process the data and to discover the
pulsars by seeking either single or periodic dispersed pulses. Of the
pulsars presented in this paper, 13 were discovered using the
Einstein@Home pipeline \citep{akc+13}, which uses spare cycles of a
global network of PCs.  The other 16 were discovered using either the
PRESTO-based pipeline \citep{ran01, lbh+15}, the ``Quicklook''
pipeline, which runs at the telescope in near-real-time
\citep{cfl+06}, or a pipeline designed to find single dispersed
pulses \citep{dcm+09}.

This paper presents some of the long-period pulsars discovered in the
survey to date. However, the survey has already resulted in the discovery of 
 binary and millisecond pulsars
\citep[e.g.,][]{lsf+06, crl+08, kac+10, kla+11, kls+15, csl+12, dfc+12,
akc+13, skl+15}, as well as the first-known repeating fast
radio burst source \citep{sch+14, ssh+16, ssh+16a}.

\section{Confirmation and Timing Observations}\label{sec:obs}

The 76-m Lovell Telescope at the Jodrell Bank Observatory, United
Kingdom, has been used both to confirm the existence of, and to
conduct follow-up timing observations of, more than half of the
pulsars detected in the survey, including the 29 pulsars reported
here.  In addition, one pulsar (J1925+1720) was also observed
initially with the Green Bank Telescope.  While some of the pulsars we
present here were discovered using the single pulse search pipeline,
all of the timing solutions presented were derived using the average
profiles of each entire observation.

Observations with the Lovell Telescope reported here mostly spanned 
the period 2011 August and 2015 January (MJD~55800$-$57050) 
and made use of a cryogenically
cooled dual-polarization receiver working in the frequency range
1350--1700 MHz. The system equivalent flux density on cold sky
is 25\,Jy. A 512-MHz-wide band was Nyquist sampled at 8-bit
resolution and channelized into 0.5-MHz-wide channels 
using a digital filterbank. To achieve a typical signal-to-noise ratio
(S/N) of $\sim6-10$, 
observation durations varied from 10 to 40 minutes. For each
pulsar the data were folded on to 1024 bins across the pulse
period in sub-integrations of 10~s using a polynomial derived
from the best ephemeris. The data were "cleaned" of radio-frequency interference by
removing sub-integrations or channels that were adversely affected. 
To form the final average profile, the dispersion delay between the
frequency channels was removed using incoherent dedispersion. 
As the initial period and dispersion measure (DM) values are uncertain, 
improved pulsar parameters were derived from the confirmation 
observations by optimising for S/N in the average profile using
a range of trial period and DM values. Once these improved parameters
were determined, subsequent analysis used times of arrival (TOAs) with
initial phase connection achieved using {\sc
Psrtime}\footnote{\url{http://www.jb.man.ac.uk/pulsar/observing/progs/psrtime.html}}
and final timing solutions were obtained using {\sc Tempo}\footnote{\url{http://tempo.sourceforge.net} or
\url{https://github.com/nanograv/tempo}}. Each of the sources had an 
observational cadence of 1--4 weeks depending on telescope availability
and typically a single TOA was derived from any given observation. 

\section{Timing}\label{sec:timing}

Timing solutions for each pulsar were found by using {\sc TEMPO} to
fit a standard timing model to the TOAs using a $\chi^2$ minimization
technique.  To incorporate the TOAs obtained with the Green Bank
Telescope for PSR J1925+1720 a time offset was computed between the
Jodrell Bank and Green Bank TOAs. Solar system and time transfer made
use of the JPL DE405 solar system ephemeris \citep{sta98b} and the
TT(BIPM) time scale\footnote{\url{ftp://tai.bipm.org/TFG/TT(BIPM)/}}
respectively.  In all cases the standard model included the rotation
period and period derivative, $P$ and $\dot{P}$, the R.A. (J2000)
$\alpha$ and Decl. (J2000) $\delta$, and the DM. As discussed below,
higher order frequency derivatives, binary parameters and glitch
parameters were also needed to adequately model some of the
pulsars. The resultant parameters and their associated 95\% confidence
interval uncertainties are given in
Table~\ref{table:measured}. Although some profiles do show scattering
(\S\ref{sec:profiles}), no correction to the DM values for those
pulsars has been made.

\begin{figure*}[p]
\begin{center}
\includegraphics[width=5.0in]{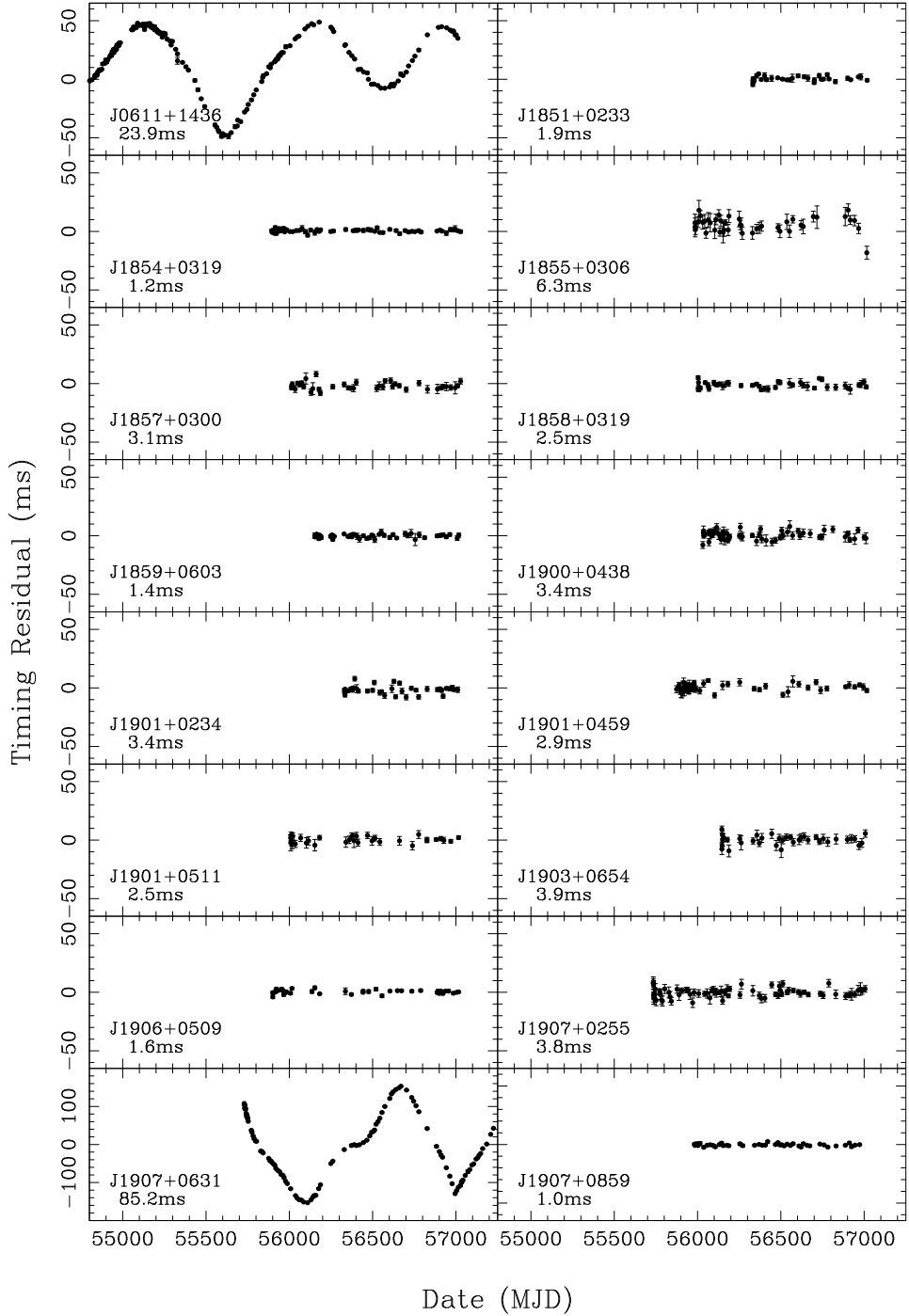}
\caption{\label{fig:resid1}Residual pulse arrival times of all pulsars
in this paper are given in Figures~\ref{fig:resid1}
and \ref{fig:resid2} to indicate the levels of timing noise.  For
each pulsar, the residual plot was made by performing a timing fit for
just spin-period and spin-down rate, with the best-fit positions given
in Table \ref{table:measured} and any binary or glitch parameters
given in Tables \ref{table:1932} or \ref{table:glitches} held
fixed. The root-mean-square of the residuals is given beneath each
pulsar name.  }
\end{center}
\end{figure*}

\begin{figure*}[p]
\begin{center}
\includegraphics[width=5.0in]{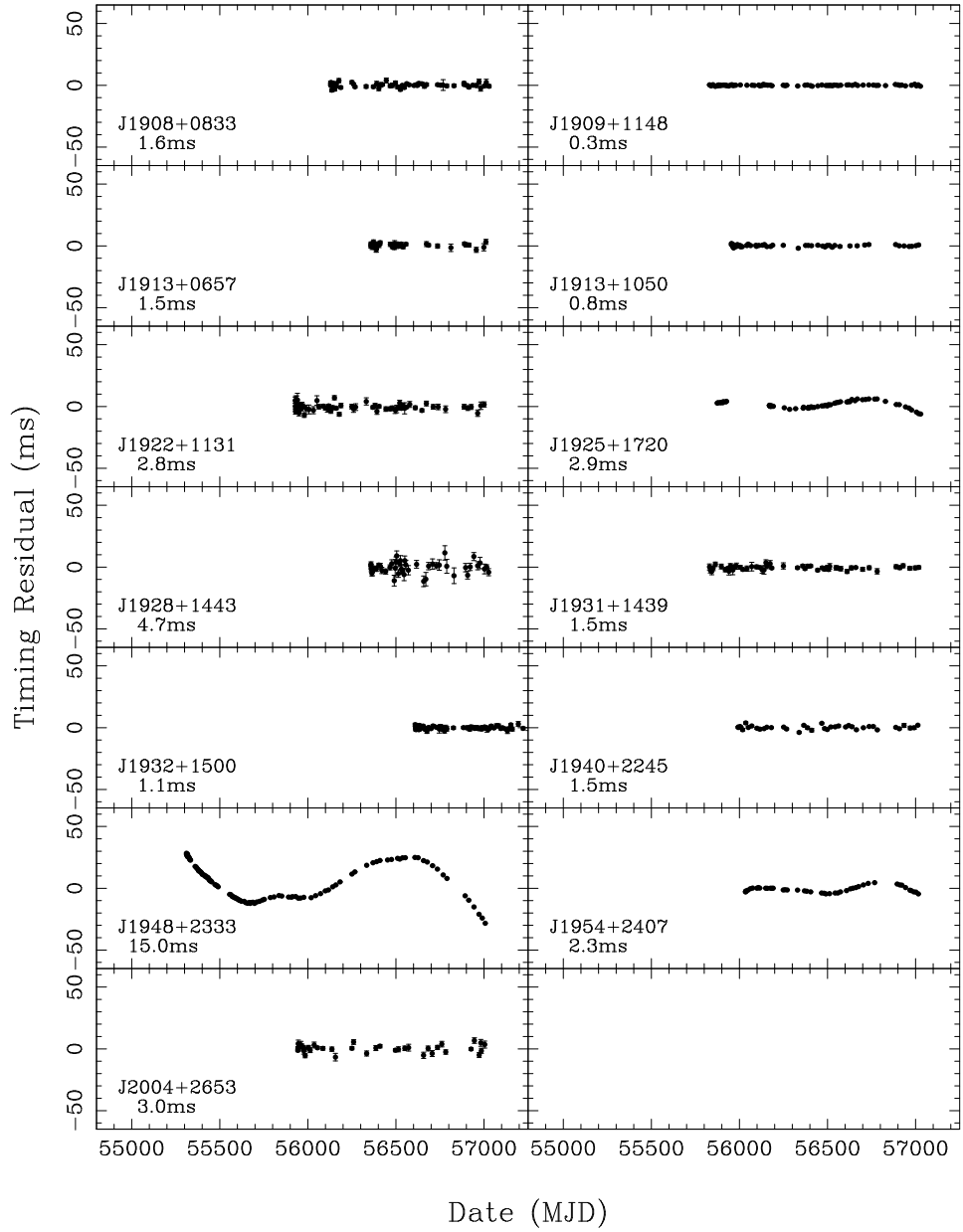}
\caption{\label{fig:resid2}See the caption for Figure \ref{fig:resid1}.}
\end{center}
\end{figure*}

The timing residuals corresponding to the difference between the
measured TOAs and the times of arrival predicted by the timing models
are shown in Figures~\ref{fig:resid1} and \ref{fig:resid2}. In the
majority of cases the residuals are "white," that is they show no
discernible structure. However, in a few cases there is significant
timing noise and higher order period derivatives were required to
"whiten" the residuals using the method described by
\citet{nab+13}. This was necessary in order to measure accurate pulsar
parameters.  The procedure ensures that the the number of fitted
derivatives $n_{\rm fit}$, listed in Table~\ref{table:measured},
results in groups of timing residuals having no evident correlation.

Several quantities, the spin-down
age, $t_{\rm s}$, surface magnetic field strength, $B$, and spin-down
energy loss rates, $\dot{E}$, derived from the timing parameters given in
Table~\ref{table:measured} are presented in
Table~\ref{table:derived}.  The table also contains the Galactic coordinates and the distance derived
from the DM and the NE2001 model \citep{cl02} for the distribution of free electrons.

\section{Pulse profiles and flux densities}\label{sec:flux}

\begin{figure*}[p]
\begin{center}
\includegraphics[width=5.0in]{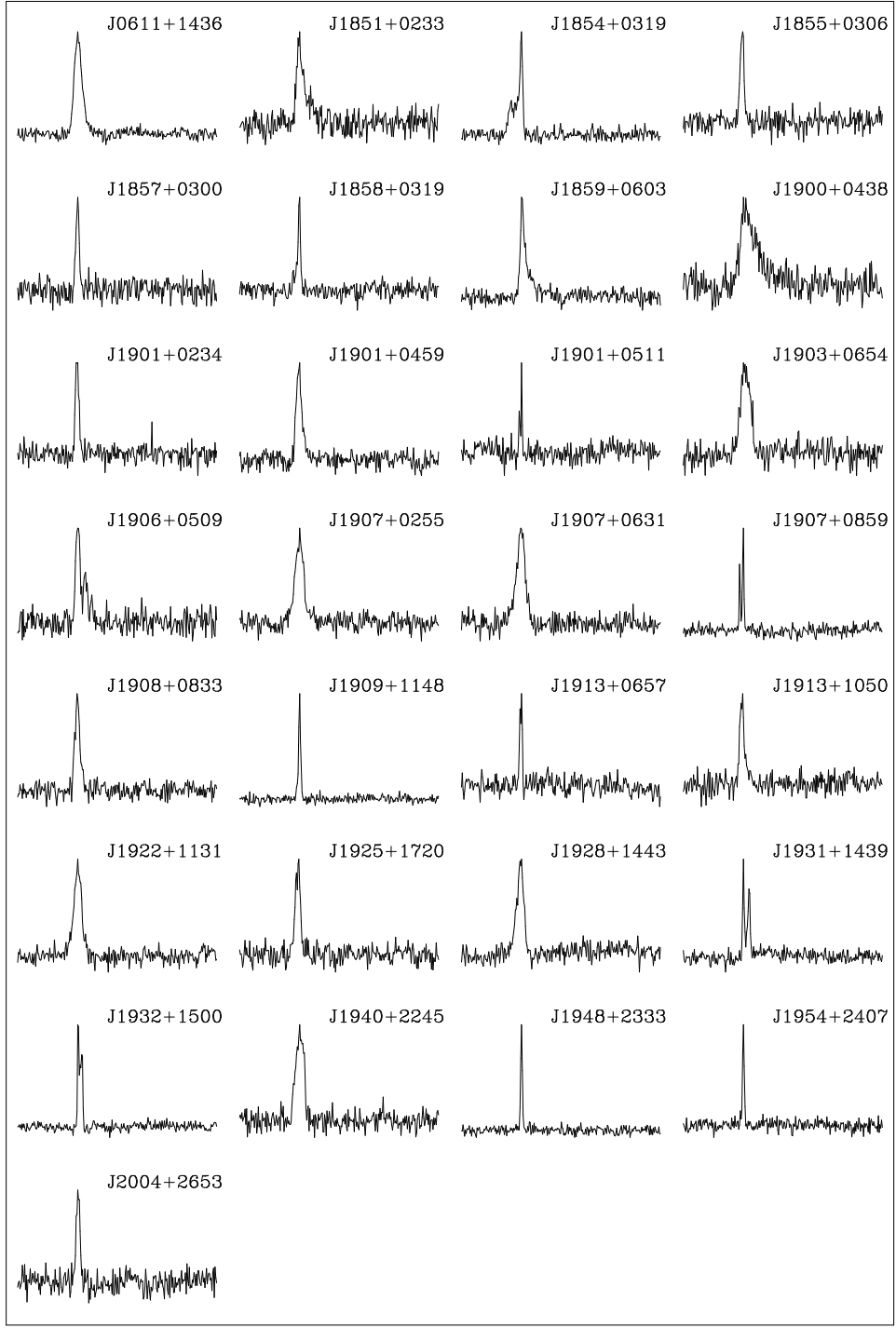}
\caption{\label{fig:profs_a} Total-intensity profiles for
each of the 29 pulsars.  The full pulse period is shown and all the profiles have been
averaged to 256 bins. All observations that contributed to the average profile used
the whitened timing solution for alignment prior to summation. In all cases there 
is no significant broadening due to the instrument or dispersion smearing. The region
around the pulse peaks is shown in more detail in Figure~\ref{fig:profs_b}. The profiles of the 29 pulsars presented in this paper and those from \citet[]{nab+13} are available from the EPN database (http://www.epta.eu.org/epndb/)}
\end{center}
\end{figure*}

\begin{figure*}[p]
\begin{center}
\includegraphics[width=5.0in]{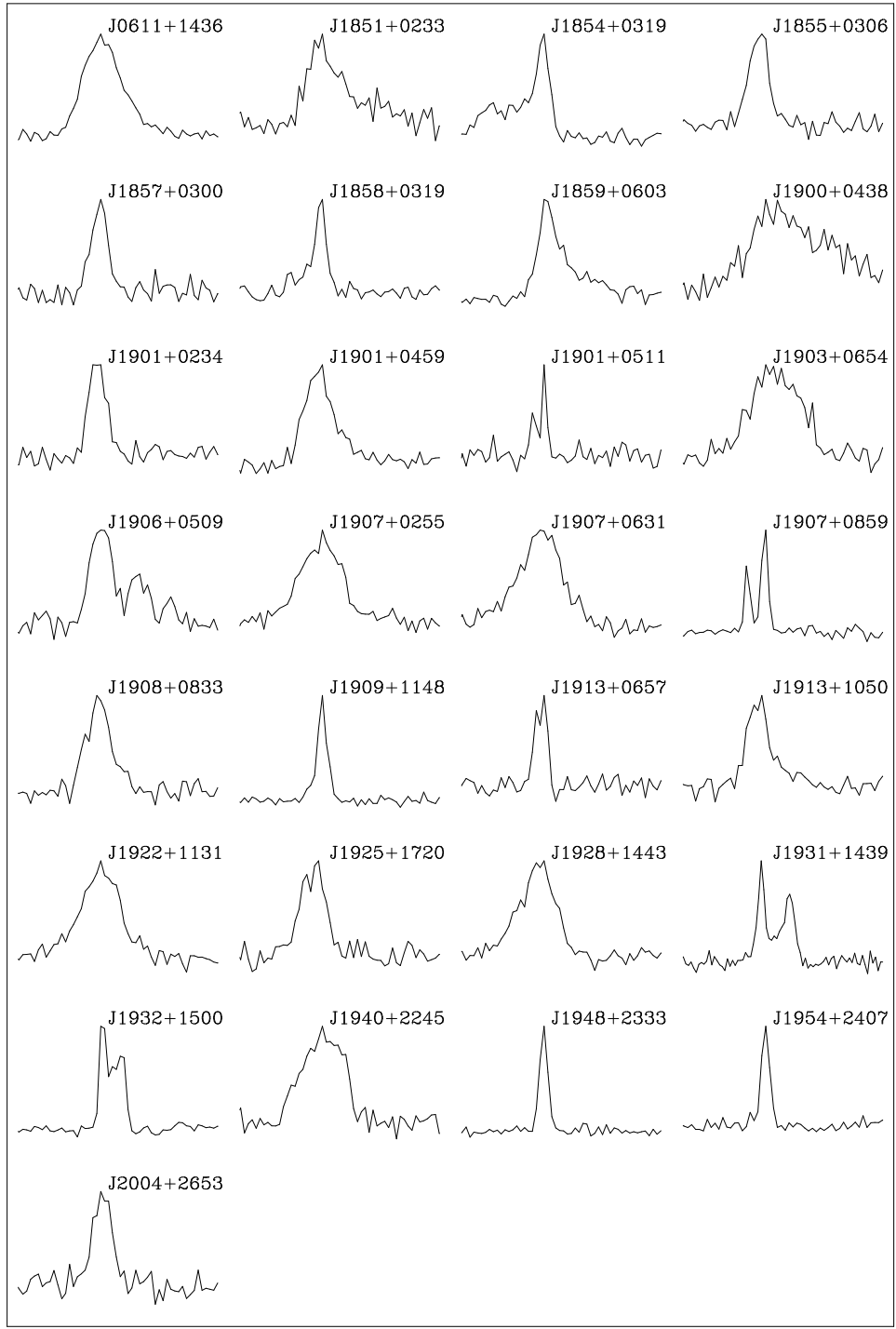}
\caption{\label{fig:profs_b}%
The same pulse profiles as presented in Figure~\ref{fig:profs_a} but
now expanded to show the 20\% of the profile around the main peak. }
\end{center}
\end{figure*}

The same Jodrell Bank timing data was used to form a long term
average pulse profile and also to determine the mean flux density of all 29 pulsars (Figures~\ref{fig:profs_a} and
\ref{fig:profs_b}). 
Only observations that resulted in a good TOA were used. The amplitude scale
 of each pulse profile was adjusted to be in units of the system noise and they were aligned using the "whitened" timing solutions (\S\ref{sec:timing}) before being summed. 

The system equivalent flux density for each pulsar position was
calculated from observations of standard continuum radio sources,
adjusted for the local sky brightness temperatures.  After appropriate
scaling, the time-averaged mean flux density $S_{1400}$ was measured
for each profile and is given in
Table~\ref{table:measured}. Luminosity estimates,
$L_{1400}=S_{1400}d^2$, are also calculated using the distances
derived from DMs using the NE2001 model \citep{cl02} and are given in
Table~\ref{table:derived}.

\subsection{Profiles and interstellar scattering}\label{sec:profiles}

The observed profiles display similar forms to those of the known
pulsar population and none exhibits a significant interpulse. The flux
density limit for any interpulse emission depends upon the
S/N ratio which varies from pulsar to pulsar and can be
judged from the profiles presented in Figure~\ref{fig:profs_a}.  This
limit is typically 5\% of the flux density of the main pulse given in
Table~\ref{table:derived}, assuming that any interpulse has a similar
pulse width to that of the main pulse. The widths of the pulse
profiles at half the peak height $W_{50}$ in milliseconds are
presented in Table~\ref{table:measured} and are also typical of the
pulsar population.

\begin{figure*}[p]
\begin{center}
\includegraphics[width=4.0in, angle=270]{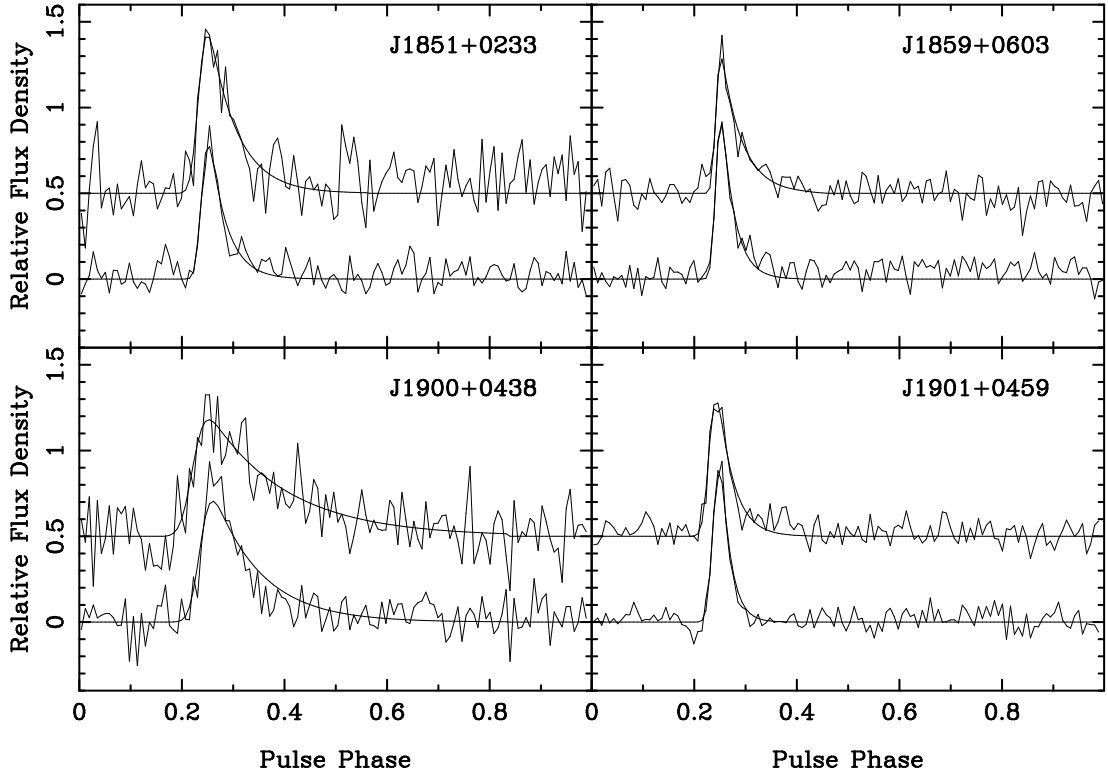}
\caption{\label{fig:scatter}
Pulse profiles from PSRs~J1851+0233, J1859+0603, J1900+0438 and J1901+0459 exhibiting 
broadening due to scattering in the interstellar medium. In
each frame, the upper and lower profiles are centered on 1450~MHz and
1650~MHz, respectively. The smooth curves are the best-fit model, see the text for details, to the displayed profiles.}
\end{center}
\end{figure*}

The relatively high DM values for these pulsars suggests that some of
them may show evidence of an asymmetric broadening of the pulse
profile with a strong frequency dependence that would be indicative of
scattering by the turbulent interstellar medium. As can be seen in
Figure~\ref{fig:scatter} this is clearly the case for PSRs J1851+0233,
J1859+0603, J1900+0438, and J1901+0459. To quantify the degree of
scattering we follow the procedure outlined in \citet{nab+13}, who
model the pulse profile as a single Gaussian component convolved with
a one-sided exponential that has a $e^{-1}$ broadening time with a
frequency dependence given by $\tau_{\rm iss}(f)=\tau_{\rm
1GHz}(f/{\rm 1\,GHz})^{-4}$. The amplitude of the Gaussian component
is allowed to vary as a function of frequency but the position and
width are held fixed. To allow for the fact that the DM used to
generate the profiles to be fitted may be biased due to distortions of
the pulse shape by the scattering, the model also includes a
$\Delta$DM fit parameter to account for this.

This model fitted the observed pulse shapes satisfactorily for all
four pulsars. The measured scattering timescales for PSRs J1851+0233,
J1859+0603, J1900+0438, and J1901+0459 in the two bands were found to
scale with observing frequency approximately as $\nu^{-4}$, confirming
the presence of interstellar multipath scattering. The pulse profiles
for the four pulsars and their fitted models are shown in
Figure~\ref{fig:scatter}.  The scattering time scales $\tau_{\rm
1GHz}$ were determined with precision of order ten percent and had
values of 0.08(1), 0.09(1), 0.29(3), and 0.12(2)~s, respectively.
Although this model fits the observed pulse profiles well, we note
that it is possible that the intrinsic pulse profiles are more complex
than the assumed Gaussian form, and this may have an unrecognized
influence on the measured values.  In order to provide truer
estimates of DM, the values given for the four pulsars in
Table~\ref{table:measured} include the fitted values of
$\Delta$DM.

In addition, PSRs~J1857+0300, J1901+0234, and J1903+0654 show some
evidence of the expected frequency scaling for scattering, although
determination of the amount of scattering was not possible due to the
low S/N ratio and/or complex profile morphology.  The
profiles of the remaining pulsars show no significant scatter
broadening.

As noted above, the broadening of the pulse profile can cause the
pulses to be delayed by a frequency-dependent amount which results in
an overestimate of the DM measured simply by optimising S/N.  Such an
effect is present even if the pulse-broadening is indiscernible.

\section{Young pulsars: PSRs J1907+0631 and J1925+1720}\label{sec:young}

PSRs J1907+0631 and J1925+1720 have the smallest characteristic ages
of the 29 pulsars presented, being around 10~kyr and 100~kyr
respectively.  Further evidence of the youth of PSR J1907+0631 is
provided by the fact that it has glitched, as discussed in
\S\ref{sec:glitches}.  The ages and locations of the pulsars in the
Galactic plane suggest that they might be associated with supernova
remnants and/or pulsar wind nebulae.  We have therefore studied the
regions around each pulsar for such structures.

\subsection{PSR~J1907+0631}

\begin{figure*}[p]
\begin{center}
\includegraphics[width=6.5in]{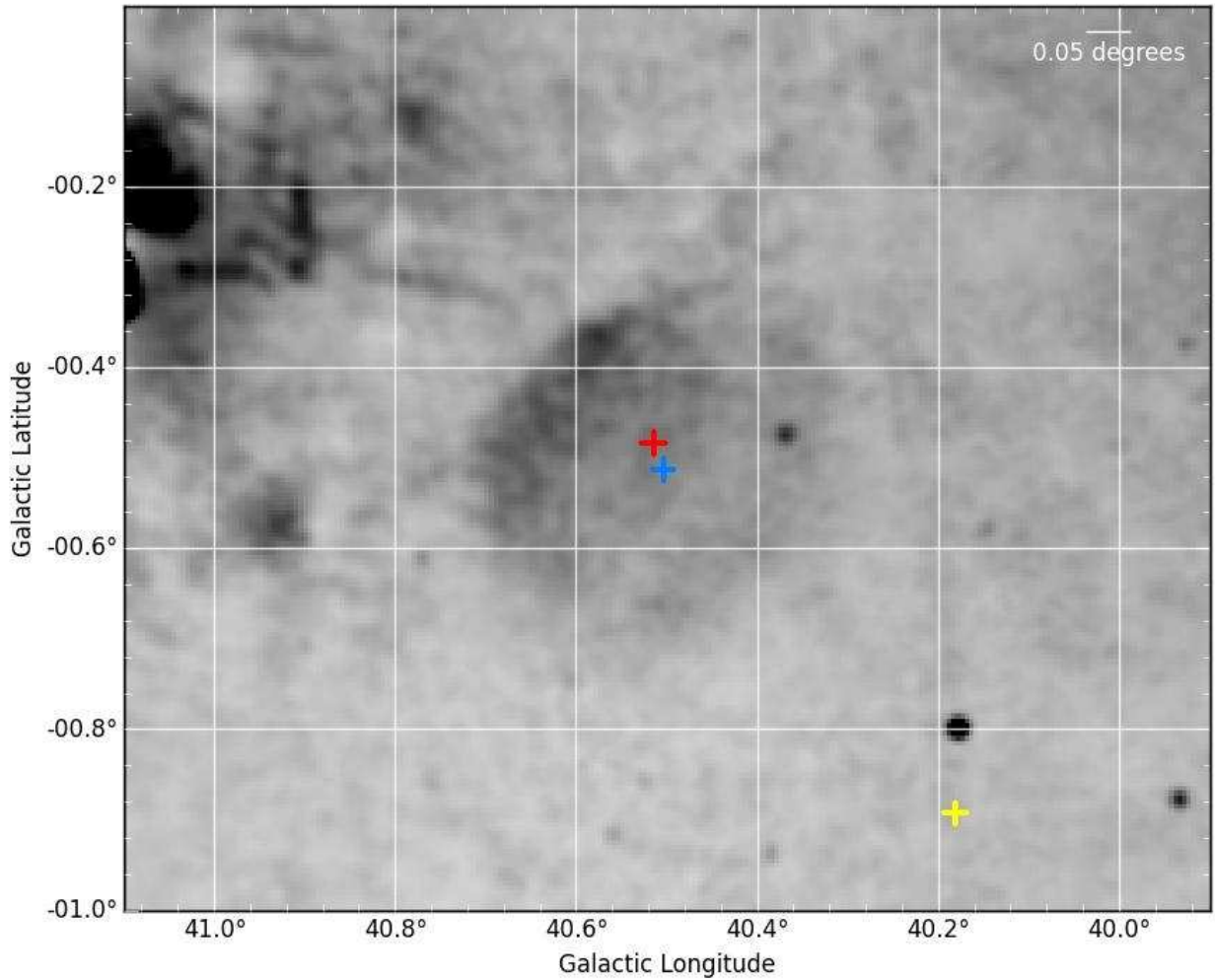}
\caption{\label{fig:1907_snr} A radio image of the region around
PSR~J1907+0631 in galactic coordinates, taken from the VGPS at
1420\,MHz \citep{std+06a}.  The red cross is
the position of the pulsar, the blue cross is the center of the
supernova remnant G40.5$-$0.5 given by Abdo et al. (2010) and the
yellow cross is the position of the gamma-ray pulsar, PSR~J1907+0602.}
\end{center}
\end{figure*}

An inspection of Green's catalog
\citep{gre14} quickly revealed that PSR~J1907+0631 is located in SNR
G40.5$-$0.5. In Figure \ref{fig:1907_snr} we show a region of the
Galactic plane taken from the VLA Galactic Plane Survey (VGPS;
\cite{std+06a}) at 1420\,MHz which shows that the pulsar (red cross)
is in fact very close to the center (blue cross; R.A. (J2000) = 19:07:08.6,
Decl. (J2000) = 06:29:53) of the remnant. We have used the position
of the center of the nebula derived by \citet{aaa+10}, which is based
on the morphology of the nebula, rather than the previous position
obtained by Langston et al. (2000)\nocite{lmd+00} which was based on
single dish radio measurements and which were biased by the bright rim
of the asymmetric remnant. It is interesting to note that there
appears to be a region of enhanced radio emission at the position of
PSR~J1907+0631 which might be from a pulsar wind nebula; further
multi-wavelength study is required to confirm this.

The age of G40.5--0.5 is often quoted as being 20--40 kyr
\citep{dsp80} which is broadly consistent with the derived
characteristic age of the pulsar. This age is based upon an estimate
of the distance to the remnant that ranges between 5.5 and 8\,kpc and
is derived from various $\Sigma-D$ relations which relate the diameter
of a remnant to its observed radio surface brightness
\citep{cc76}. Yang et al. (2006)\nocite{yzc+06} study the molecular
gas distribution around the remnant and also derive their own
$\Sigma-D$ distance of 5.3\,kpc. We note that such distances based
upon $\Sigma-D$ relationships should be treated with caution
\citep[e.g.][]{gre15}.  However, Yang et al. (2006) prefer a distance
of 3.4\,kpc which they derive from the velocity of a CO feature from
shocked gas that they associate with the remnant and the distances
from Clemens et al. (1988)\nocite{css88}. This contrasts somewhat with
the NE2001-derived distance to the pulsar of 7.9\,kpc for the observed
DM of 427\,pc\,cm$^{-3}$. However as can be seen in
Figure \ref{fig:1907_snr} and discussed in Yang et al. (2006), this is
a complex region with many molecular features, while the line of sight
to the pulsar and the remnant passes through the nearside of the
Sagittarius arm at about 3\,kpc and the far side at about 9\,kpc.

There has been recent discussion about the possibility that the nearby
gamma-ray pulsar J1907+0602 \citep{aaa+09} might be associated with
G40.5--0.5 because of its association with the nearby TeV source MGRO
J1908+06 \citep{aab+07} which extends to include the lower limb of the
remnant closest to PSR~J1907+0602 (see Figure 1 of Aliu et
al. (2014)\nocite{aaa+14}). Abdo et al. (2010) \nocite{aaa+10} found
that this pulsar (yellow cross in Figure \ref{fig:1907_snr}) lies in
the extended emission seen also with the High Energy Stereoscopic
System (HESS; Aharonian et al. 2009)\nocite{aaa+09b} and the Very
Energetic Radiation Imaging Telescope Array System (VERITAS; Ward
2008)\nocite{war08}. They noted, however, that the association with
G40.5--0.5 seems very unlikely as it would require an extremely high
transverse velocity and should have resulted in a bow-shock shaped
nebula near the pulsar which was not evident in their ${\it Chandra}$
X-ray observations.  This is consistent with our belief that
PSR~J1907+0631 is in fact the pulsar that was formed in the G40.5--0.5
supernova event.

The position identified by Abdo et al. (2010) as the center of the
remnant seems to be based on the apparently symmetric shape of the
nebula. However, they do not quote a position error and the
clearly asymmetric intensity distribution of the 1400\,MHz radio
emission could argue for an offset in the location of the progenitor
star from their choice of position. Despite this, we use their
position for the purpose of deriving a separation of PSR~J1907+0631
from the center of the remnant, which we find to be 1.8
arcminutes. Assuming that the remnant distance is 3.4\,kpc and an age
for the pulsar of 11\,kyr, this separation requires the pulsar to have
an unremarkable transverse velocity of 155\,km\,s$^{-1}$. This would
be approximately doubled, and still unremarkable, if the DM-based
distance of 7.9 kpc for the pulsar were the true distance.  Deeper and higher
resolution radio images of the remnant could confirm whether the radio
emission close to the pulsar position is indeed a plerion and if it
has a bow-shock-like form. The system may then be similar to
PSR~B1853+01 in W44 \citep{fggd96} and we can learn more about the
velocity and wind of the pulsar from radio and X-ray \citep{pks02}
studies. 

PSR~J1907+0631 falls within the field of view of an 
\textit{XMM$-$Newton} observation which targeted MGRO~1908+06 (ObsID
0553640801). There is no X-ray source detected at the position of the
pulsar.  Based on the measured DM and DM-derived distance of 7.9 kpc,
we estimate the hydrogen column density along the line of sight to
PSR~J1907+0631 to be $N_{\rm H} \approx 1.3\times10^{22}$ cm$^{-2}$
using the empirical relation from \citet{hnk13}. Given the
non-detection, from the more sensitive EPIC pn data we place a
3$\sigma$ upper limit on the unabsorbed X-ray luminosity of $L_X \sim
4\times 10^{32}$ erg s$^{-1}$ in the 0.3--10 keV band. The implied
spin-down to X-ray conversion efficiency ($L_X/\dot{E}$) of PSR
J1907+0631 is $\lesssim$$8 \times 10^{-4}$, which is typical for young
pulsars with comparable $\dot{E}$ \citep[see, e.g.,][]{kdpg12}.

\subsection{PSR~J1925+1720}

PSR~J1925+1720 is not only youthful but has a sufficiently high
$\dot{E}$ that, despite a relatively large NE2001-derived distance
of 6.9\,kpc, places its $\dot{E}/{d}^2 =
2.1\times10^{34}$\,erg\,kpc$^{-2}$\,s$^{-1}$ in the top 10\% of all
known pulsars. We have therefore sought the presence of any
high-energy nebula that might be related to the pulsar.

Inspection of the \textit{Fermi} Large Area Telescope 4-year Point
Source Catalog \citep[3FGL; see][]{aaa+15} reveals that the radio
position of PSR~J1925+1720 places it 7.4$'$ from the high-energy
$\gamma$-ray source 3FGL J1925.4+1727, at R.A. (J2000) = $19^{\rm h}25^{\rm
m}29^{\rm s}.6$ Decl. (J2000) = $+17^{\circ}27'48''$, with a 95\% error ellipse of
size $6'\times5.4'$, and position angle $-15.6^{\circ}$. Although the
pulsar falls outside of the error ellipse and the spectrum of
3FGL~J1925.4+1727 is not obviously pulsar-like, our investigation
indicates that this is caused by the presence of an additional
unmodeled source, which appears to be a common occurrence in the
Galactic plane \cite[see, e.g.,][for the case of
PSR~J1906+0722]{cpw+15}. 

We note that at X-ray energies, PSR J1925+1720 has been observed with
\textit{Swift} XRT for a total effective exposure of 3.4 ks. No
sources are apparent in the resulting X-ray image. Taking DM = 223.3
pc cm$^{-3}$ and $d=6.9$ kpc for the pulsar, we estimate $N_{\rm H}
\approx 7 \times 10^{21}$ cm$^{-2}$ as per \citet{hnk13}. From the
available X-ray data, we then obtain a 3$\sigma$ upper limit of $L_X
\sim 2\times 10^{33}$ erg s$^{-1}$ (0.3--10 keV) on the unabsorbed
X-ray luminosity. The implied $L_X/\dot{E} \lesssim 2 \times 10^{-3}$
is consistent with $L_X/\dot{E}= 10^{-4}- 10^{-3}$ observed for
analogous energetic pulsars \citep[see Figure 2 in][]{kdpg12}.

A detailed analysis of the $\gamma$-ray emission from PSR~J1925+1720
will be presented in a follow-up publication (S.~Bogdanov et al. 2016,
in preparation).

\section{A Long-period pulsar in a 199-day binary orbit: PSR J1932+1500}\label{sec:binary}

Shortly after its discovery, timing observations of this 1.86-second
pulsar with the Lovell Telescope indicated that the period was
changing rapidly, presumably due to the Doppler effect arising from its
motion in a binary orbit with a companion star.  Subsequent observations
confirmed this, and showed it to be in an essentially circular orbit.
The parameters resulting from a full fit to the timing data of this
pulsar are given in Table~\ref{table:1932}.

The orbital period, eccentricity and spin characteristics of the
PSR~J1932+1500 binary system are very similar to those of a previously
known pulsar, PSR~J1822$-$0848 \citep{lfl+06}.  The spin period and
period derivatives of both pulsars indicate that they were only mildly
recycled - in the $P$ - $\dot{P}$ diagram they are located in the
region where most so-called ``normal'' pulsars lie. However, as
described below they must have experienced some accretion and spin-up.

The orbital eccentricities of these systems (0.0289 and 0.0589 for
PSRs J1932+1500 and J1822$-$0848 respectively) and the estimated
companion masses ($\sim 0.30\, {\rm M}_{\odot}$ for both systems) provide
important clues about their previous evolution.  Both eccentricities
are much smaller than for truly unrecycled, recently formed pulsars in
binary systems, such as PSRs J0045$-$7319, B1259$-$63, J1638$-$4725,
J1740$-$3052, and J2032+4127
\citep{mmh+91,jlm+92,lfl+06,sml+01,lsk+15}, where the pulsars are very
young with no signs of significant prior accretion and their massive
(from $\sim 3$ to $\sim\, 12 {\rm M}_{\odot}$) companion stars have not yet
entered their giant phases. In these cases, the orbital eccentricities
range from 0.6 to 0.96, implying that there has been no time yet for
tidal circularization. That will eventually happen when the companion
stars evolve into giant stars: their sizes will then be comparable to
the minimum orbital separation, and tidal effects will then be very
important.

Therefore, it is very likely that the orbits of PSRs~J1932+1500 and
J1822$-$0848 have been tidally circularized by their companion in a
previous giant stage of their evolution. This was not merely a
dynamical interaction -- it probably affected the pulsars as well:
their $P$ and $\dot{P}$ put them just below the spin period - magnetic
field equilibrium line of \citet{pr72} for accretion at the Eddington
rate, i.e. their spin characteristics are consistent with mild
accretion by mass transfer from the envelope of the companion.
Furthermore, their minimum masses follow the Tauris \& Savonije
(1999)\nocite{ts99a} relation for their orbital periods (see Fig. 4 in
that paper).

\begin{figure*}[p]
\begin{center}
\includegraphics[width=4.0in,angle=270.0]{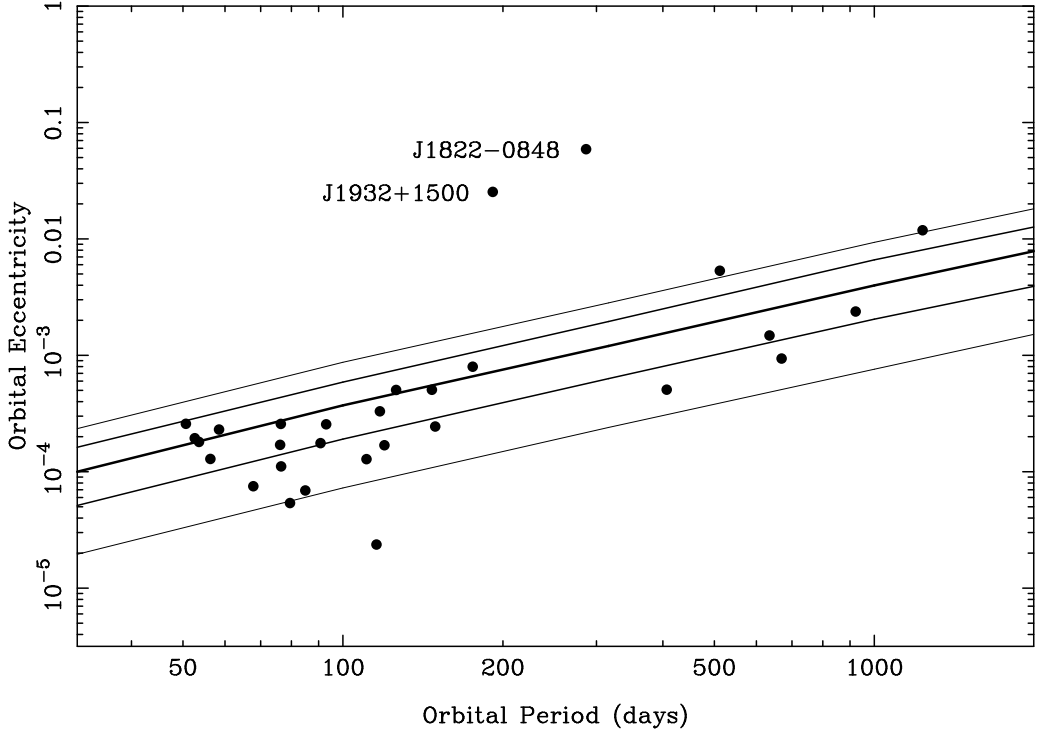}
\caption{\label{fig:pb_ecc} The anomalous eccentricities of
PSRs~J1932+1500 and PSR~1822$-$0848, compared with those of other
long-period, mildly recycled pulsars and the predictions of
\citet{phi92}. This diagram includes all known Galactic binary pulsars
with orbital period greater than 50 days and minimum companion mass
less than 0.5~M$_\odot$ (Table \ref{table:wide-orbits}). The central
curve represents the median eccentricity predicted by the convective
fluctuation-dissipation theory of \citet{phi92}. The adjacent pairs of
lines are predicted to contain 68\% and 95\% of the resultant
eccentricities. Excluding PSRs~J1932+1500 and PSR~1822$-$0848, the
observed occupancies of the two ranges are 56\% and 85\% respectively
and are reasonably consistent with the theory.}
\end{center}
\end{figure*}

There are quite a few other moderately recycled pulsars with longer
orbital periods that we list in Table~\ref{table:wide-orbits}, the
longest being PSR~B0820+02 (Manchester et al. 1980)\nocite{mncl80},
with $P_b = 1232$ days, where the companion is a 0.6-0.7 M$_{\odot}$
WD \citep{kcr92}.  Many of these binaries (notably PSRs J0214+5222,
J2016+1947, J0407+1607, J1711$-$4322, and J1840$-$0643) have much
shorter spin periods, much smaller B-fields and smaller orbital
eccentricities, which places them more in line with the predictions of
the convective fluctuation-dissipation theory of \citet{phi92}. This
is not the case for PSRs~J1932+1500 and J1822$-$0848: although small,
their eccentricities are $\sim$1.5 orders of magnitude larger than the
prediction of Phinney (1992), as can be seen in Figure
\ref{fig:pb_ecc}.

The reasons for these differences are not clear. Even if
PSRs~J1932+1500 and J1822$-$0848 had much stronger B-fields to start
with, or perhaps the accretion and related B-field burial was not as
efficient, that would not explain why the orbital eccentricities are
so much higher than for the other recycled pulsars with wide orbits.

A possible explanation for the characteristics of PSR~J1822$-$0848 was
advanced by \citet{tlk12}. In their scenario, PSR~J1822$-$0848 was
initially a non-recycled pulsar in a wide, eccentric orbit. When its
companion reached the asymptotic giant branch stage, the system
experienced weak spiral-in (and relatively inefficient orbital
circularization) from an almost unbound common envelope.  This might
explain its spin characteristics (very weak recycling, if at all) and
its current orbital period and eccentricity, both of which are much
smaller than for the unrecycled pulsar - O-B star systems.  This
hypothesis would also explain the characteristics of PSR~J1932+1500.

The optical characteristics of the companion of PSR~J1932+1500
represent an important constraint on the nature and evolution of this
system.  There is a faint candidate in the SDSS \citep{aaa+15a} at
coordinates $\alpha \,= 19^{\rm h}32^{\rm m}46^{\rm s}.332,\, \delta\,
= \, +15^{\circ}00'23''.7$.  The angular offset from the pulsar
position is 1.3$''$; this is significantly larger than the astrometric
uncertainties of the catalog (0.15 arcseconds,
http://classic.sdss.org/dr7/instruments/technicalPaper/index.html) and
more than 3 times the 1-$\sigma$ positional uncertainty of the pulsar
(0.4$''$).  This source is not detectable in the 2 micron All-Sky
Survey \citep[2MASS, ][]{scs+06}. The significant offset leads us to
conclude that the SDSS source is unrelated to the pulsar; its
proximity is the result of the large density of stars detected by SDSS
in this region of the sky. This non-detection of the companion star
suggests that it is a compact object, as predicted by the hypothesis
of \citet{tlk12}.

\section{Large glitches in PSRs J0611+1436 and J1907+0631}\label{sec:glitches}

\begin{figure*}[p]
\begin{center}
\includegraphics[width=4.0in,angle=270.0]{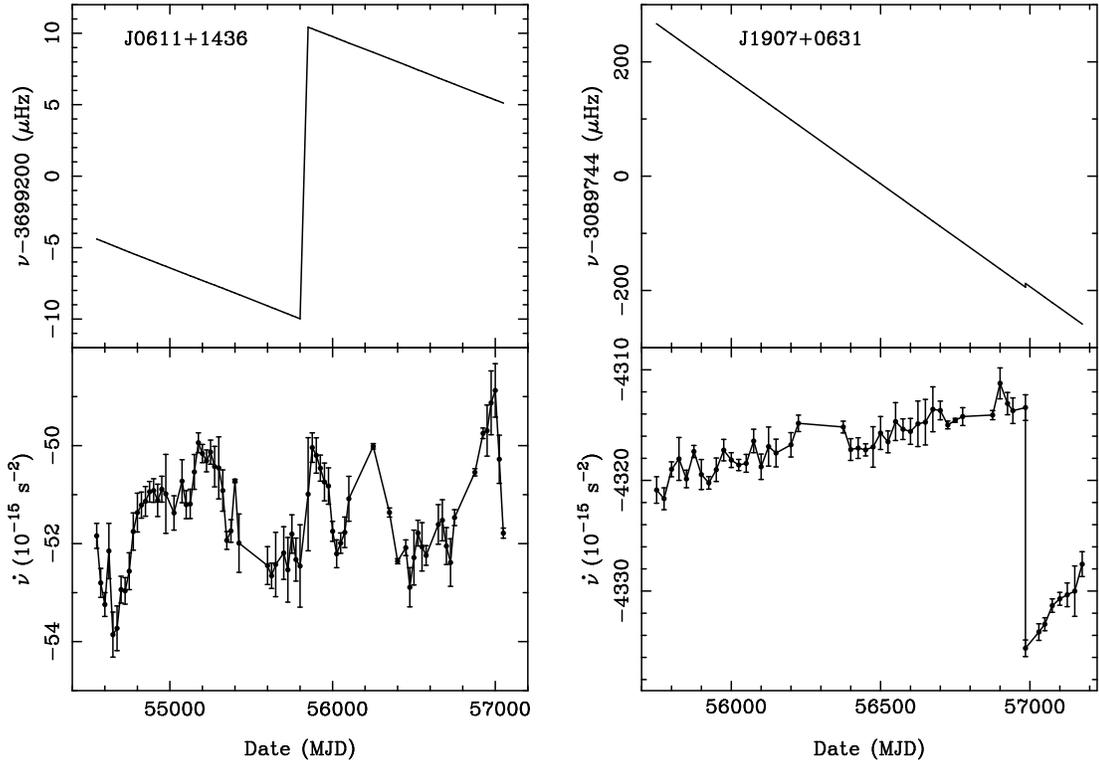}
\caption{\label{fig:glitches_f0f1} The large glitches in
PSR~J0611+1436 and PSR~1907+0631. Top: the evolution of the rotation
frequency $\nu$. Bottom: the evolution of the frequency derivative
$\dot\nu$.  Note that in PSR~J0611+1436, the glitch reverses
approximately 12 years of normal spin down, compared with 17 days in
PSR~1907+0631.}
\end{center}
\end{figure*}

Timing observations of these two pulsars, with periods of around
0.3~s, showed that they both suffered large glitches in their rotation
rates during the timing program, with fractional increases of
$5.5\times10^{-6}$ and $2.1\times10^{-6}$, respectively.  However, the
evolution of the rotational frequencies and their first derivatives
shown in Figure \ref{fig:glitches_f0f1}, and the relevant parameters
summarized in Table~\ref{table:glitches}, reveal that they are very
different events.

PSR~J1907+0631, with a characteristic age of 11~kyr, is a young
Vela-like pulsar, still surrounded by a visible SNR and possibly a plerionic
nebula (\S\ref{sec:young}).  The observed glitch in this pulsar
(Figure \ref{fig:glitches_f0f1}) is also typical of glitches in the
Vela pulsar and similar pulsars such as B1727$-$33, B1800$-$21, and
B1823$-$13, having characteristic ages of 10-20~kyr
\citep{lsg00,elsk11}. Such glitches are characterized by fractional
increases in spin-rate of a few parts per million, followed by a
short-term recovery which is completed on a timescale of typically two
hundred days, leading into a period of approximately constant
$\ddot{\nu}$ until the next glitch occurs.  During this last phase,
the braking index, $n=\nu\ddot{\nu}/\dot{\nu}^2$, takes a large value
of a few 10s, and is about 10 in the case of J1907+0631.  This
compares with the expected value of $n=3$ for normal magnetic braking
of a rigid neutron star with a constant dipolar magnetic field. Note
that the step in rotational frequency at the glitch is seen to be a
minor perturbation in the systemic spin-down of the pulsar, and
``undoes'' only 17 days of spin-down.

On the other hand, it is clear from Figure~\ref{fig:glitches_f0f1}
that PSR~J0611+1436, with a characteristic age of 1.1 Myr, shows
essentially no short-term transient behavior following the even
larger glitch in frequency, and will take $\sim$12 years to slow down
to the pre-glitch rotation rate. This is an extreme recovery time
which is not surpassed by glitches in any other normal pulsar, the
next largest being 4.9 years in PSR~B0355+54 \citep{lyn87} and 4.5 years in
B1535$-$56 \citep{jml+95}.  Figure~\ref{fig:glitches_f0f1} also shows
that the interglitch rotation rate is dominated by stochastic
variations due to timing noise which is prevalent in pulsars of
intermediate age \citep{hlk10}.

The different recovery behavior of glitches between young and
intermediate-age pulsars has been pointed out before by \citet{lsg00}
and studied more recently by \citet{elsk11}.  It seems that the
properties may be related to the decreasing amount of superfluid
neutrons as pulsars age.  We also note that glitch recoveries in
magnetars also show great diversity \citep{dk14} and may also involve
changes in the neutron star magnetospheres.  In this context, it is
notable that PSR~J1907+0631 has a large surface magnetic field of just
over $10^{13}$ G, which is greater than that of 98\% of the known
radio pulsars.

\section{Conclusions}\label{sec:conclusions}

The deep search of the Galactic plan undertaken by the PALFA survey
has revealed another 29 normal pulsars. The excellent sensitivity of
the survey and high frequency resolution of the two data acquisition
systems used means that these pulsars have large DMs
and are located at a median distance of 7.0\,kpc. This value is very
similar to the 7.1\,kpc of the 35 slow PALFA pulsars presented by
\citet[]{nab+13}, who already noted that this was much larger than
the median distance for the known Galactic population of 4.2\,kpc
(calculated using the ATNF pulsar catalog \citep{mhth05}).

The profiles presented here are indistinct from the normal pulsar
population, although a handful do show evidence for scattering and
combined with the DM can be added to the PALFA sample
to aid in modeling the ionized interstellar medium in the survey
region \citep{nab+13}. This could be enhanced by further observations
of this sample at lower frequencies.  The combined total of 64 pulsars
presented in this paper and in \citet[]{nab+13} are just the first of
the ``normal pulsars'' discovered in this survey.  The ongoing PALFA
survey has to date discovered more than 169 pulsars and, as they probe
a unique section of the pulsar population in our Galaxy, they will
provide an important statistical sample for modeling the Galactic
pulsar population and the distribution of the magneto-ionic medium.

As well as adding to the population as a whole, the sample presented
here includes two new young pulsars which both show evidence of
interacting with the local medium. PSR J1907+0631 lies close to the
center of SNR G40.5--0.5 and thus likely solves the issue of which of
the many neutron stars in the area are associated with the
remnant. This not only secures another SNR-pulsar association for
further detailed study, it clarifies some of the likely properties of
pulsars like the gamma-ray source PSR J1907+0602 which had a claim of
association. It may also help understand the origin of the TeV
emission seen in this region. The youthful and high-$\dot{E}$ pulsar
J1925+1720 might be a gamma-ray source or be powering a high-energy
nebula. PSR~J1907+0631 also exhibited a glitch, as did J0611+1436,
and, although they were both large events, they exhibited very
different period and period-derivative evolutions which not only mark
them as interesting sources to continue to monitor but also provide
important additional data for our understanding of the nature of
glitches.

The long period binary pulsar J1932+1500 appears to form a group of
two with PSR J1822--0848 that have formed in a system which initially
had a wide, eccentric orbit, that underwent only modest spiral in and
thus relatively inefficient orbital circularization and likely has a
compact companion. This source highlights the wide range of possible
outcomes that are possible in the evolution of binary systems which later
form at least one neutron star.

\acknowledgements 

The Arecibo Observatory is operated by SRI International under a
cooperative agreement with the National Science Foundation
(AST-1100968), and in alliance with Ana G. M\'endez-Universidad
Metropolitana, and the Universities Space Research Association.
Pulsar research at Jodrell Bank and access to the Lovell Telescope is
supported by a Consolidated Grant from the UK's Science and Technology
Facilities Council. This work was supported by the Max Planck
Gesellschaft and by NSF grants 1104902, 1105572, and 1148523. P.C.C.F., P.L.,
and L.G.S. gratefully acknowledge financial support by the European
Research Council for the ERC Starting Grant BEACON under contract
no. 279702. J.vL. acknowledges funding from the European Research
Council under the European Union's Seventh Framework Programme
(FP/2007-2013) / ERC Grant Agreement no. 617199. J.S.D. was supported
by the NASA Fermi Guest Investigator program and by the Chief of Naval
Research. J.W.T.H. acknowledges funding from an NWO Vidi fellowship
and from the European Research Council under the European Union's
Seventh Framework Programme (FP/2007-2013) / ERC Starting Grant
agreement no. 337062 (``DRAGNET'').  Pulsar research at UBC is supported
by an NSERC Discovery Grant and by the Canadian Institute for Advanced
Research. V.M.K. receives support from an NSERC Discovery Grant, an
Accelerator Supplement and from the Gerhard Herzberg Award, an
R. Howard Webster Foundation Fellowship from the Canadian Institute
for Advanced Study, the Canada Research Chairs Program, and the Lorne
Trottier Chair in Astrophysics and Cosmology. The National Radio Astronomy
Observatory is a facility of the National Science Foundation operated
under cooperative agreement by Associated Universities, Inc..

We thank Thomas Tauris for the interesting discussions on the
evolution of these pulsars.  This research has made use of NASA's
Astrophysics Data System Bibliographic Services.
We also thank all Einstein@Home volunteers, especially those whose
computers found the pulsars with the highest statistical
significance. PSR J1855+0306: Jeroen Moetwil, Flagstaff, Arizona, USA
and Robert E. Inman Jr., Virginia Beach, Virginia, USA. PSR
J1857+0300: ``Philemon1752'' and ``edgen.'' PSR J1858+0319: Philipp
Khlitz, Berlin, Germany and ``Termit.'' PSR J1900+0438: Robert
D. Burbeck, Ilkeston, UK and Harald Buchholz, Springfield, Virginia,
USA. PSR J1901+0511: John A. Lorimer, Canville, California, USA and
Ugur Munir Kir, Istanbul, Turkey. PSR J1903+0654: Paul Frei, Altnau,
Switzerland and Jyrki Ojala, Turku, Finland. PSR J1906+0509: Piotr
Kamiski, Warsaw (Poland) and ``zebo-the-fat.'' PSR J1907+0859: Naohiro
Goto, Ichikawa, Japan and Thomas Herdtle, St. Paul, Minnesota,
USA. PSR J1908+0833: Josef Hahn, Neuss, Germany and Charles Adams,
Alamogordo, New Mexico, USA. PSR J1913+1050: Zsolt Szvoboda,
Szentendre, Hungary and ``Rensk.'' PSR J1922+1131: Edvin Grabar, Pula,
Croatia and Peter van der Spoel, Utrecht, Netherlands. PSR J1954+2407:
Andrew Fullford, Dallas, Texas, USA and Pavlo Ovchinnikov,
Dnipropetrovsk, Ukraine. PSR J2004+2653: ``Cauche Nathanael'' and Robert
La Plante, Baltimore, Maryland, USA.


\begin{deluxetable}{cllclllcrrr}
\tabletypesize{\scriptsize}
\tablewidth{0pt}
\tablecaption{\label{table:measured}Measured Parameters of 29 Pulsars\tablenotemark{a}}
\rotate
\tablehead{%
PSR & \multicolumn{1}{c}{$\alpha$} &
\multicolumn{1}{c}{$\delta$} & \multicolumn{1}{c}{Epoch} &
\multicolumn{1}{c}{$P$} & \multicolumn{1}{c}{$\dot{P}$}   &
\multicolumn{1}{c}{DM} & $n_{\rm fit}$ & $\sigma_{\rm res}$ & $S_{1400}$ &
$W_{50}$ 
\\
    & \multicolumn{1}{c}{(J2000) } & 
\multicolumn{1}{c}{(J2000) } & \multicolumn{1}{c}{(MJD)       } & 
\multicolumn{1}{c}{(s)} & \multicolumn{1}{c}{($10^{-15}$)}  &
\multicolumn{1}{c}{   (pc\,cm$^{-3}$)} &  & (ms) & (mJy) & (ms)
}
\startdata
J0611+1436  & 06:11:18.649(13)  & +14:36:52(4)      & 55818 & 0.27032946262(3)    & 3.997(3)      & 45.7(7)   & 9  &   1.2 &  1.10 &  11.6 \\ 
J1851+0233  & 18:51:09.130(19)  & +02:33:46.4(9)    & 56700 & 0.344018308374(12)  & 2.1844(13)    & 606(4)\tablenotemark{b}    & 1  &   1.8 &  0.08 &  13.5 \\ 
J1854+0319  & 18:54:00.110(12)  & +03:19:12.8(5)    & 56700 & 0.628540823296(15)  & 0.0551(6)     & 480.2(13) & 1  &   1.0 &  0.17 &  10.4 \\ 
J1855+0306  & 18:55:38.24(7)    & +03:06:22(3)      & 56700 & 1.63356517930(18)   & 6.971(8)      & 634(6)    & 2  &   5.0 &  0.06 &  35.9 \\ 
J1857+0300  & 18:57:16.91(3)    & +03:00:26.0(13)   & 56700 & 0.77267804332(4)    & 2.6489(16)    & 691(4)    & 1  &   2.8 &  0.05 &  14.0 \\ [2pt]
J1858+0319  & 18:58:40.88(3)    & +03:19:21.5(10)   & 56700 & 0.86744387855(4)    & 0.1025(15)    & 284(3)    & 1  &   2.7 &  0.06 &  10.6 \\ 
J1859+0603  & 18:59:42.131(16)  & +06:03:54.5(5)    & 56700 & 0.508561079708(17)  & 1.5895(10)    & 378.6(20)\tablenotemark{b} & 1  &   1.1 &  0.16 &  13.8 \\ 
J1900+0438  & 19:00:13.35(3)    & +04:38:46.8(11)   & 56700 & 0.312314406456(18)  & 3.2304(8)     & 627(6)\tablenotemark{b}    & 1  &   3.0 &  0.12 &  26.0 \\ 
J1901+0234  & 19:01:26.93(3)    & +02:34:51.4(10)   & 56700 & 0.88524028123(4)    & 23.034(4)     & 404(3)    & 1  &   2.9 &  0.14 &  20.4 \\ 
J1901+0459  & 19:01:17.47(4)    & +04:59:06.8(12)   & 56700 & 0.87704381015(6)    & 15.6817(20)   & 1108(4)\tablenotemark{b}   & 1  &   2.9 &  0.12 &  28.0 \\ [2pt]
J1901+0511  & 19:01:42.91(5)    & +05:11:00.2(15)   & 56700 & 4.6003689902(4)     & 25.338(15)    & 410(7)    & 1  &   2.0 &  0.05 &  23.5 \\ 
J1903+0654  & 19:03:55.28(5)    & +06:54:39.4(16)   & 56700 & 0.79123225301(7)    & 10.595(4)     & 329(7)    & 1  &   3.3 &  0.11 &  56.2 \\ 
J1906+0509  & 19:06:56.497(16)  & +05:09:35.6(6)    & 56406 & 0.397589683040(5)   & 5.2152(4)     & 99.5(19)  & 1  &   1.3 &  0.07 &  21.2 \\ 
J1907+0255  & 19:07:17.90(4)    & +02:55:02.7(10)   & 56700 & 0.61876063644(3)    & 0.2230(9)     & 257(4)    & 1  &   3.3 &  0.14 &  30.8 \\ 
J1907+0631  & 19:07:03.816(17)  & +06:31:18.9(6)    & 56985 & 0.323648024490(19)  & 452.1551(12)  & 428.6(18) & 4  &   2.0 &  0.25 &  14.8 \\ [2pt]
J1907+0859  & 19:07:01.913(10)  & +08:59:43.8(4)    & 56700 & 1.52704227084(3)    & 5.5446(12)    & 190(3)    & 1  &   0.9 &  0.07 &  35.6 \\ 
J1908+0833  & 19:08:20.960(16)  & +08:33:31.5(5)    & 56700 & 0.512110722994(12)  & 1.9872(7)     & 700.1(20) & 1  &   1.4 &  0.20 &  14.6 \\ 
J1909+1148  & 19:09:31.379(3)   & +11:48:59.86(8)   & 56700 & 0.448945466859(3)   & 0.07250(8)    & 199.8(5)  & 1  &   0.3 &  0.09 &   4.6 \\ 
J1913+0657  & 19:13:32.92(4)    & +06:57:24.3(10)   & 56700 & 1.25718110369(7)    & 2.829(7)      & 142(3)    & 1  &   1.3 &  0.05 &  19.4 \\ 
J1913+1050  & 19:13:35.364(9)   & +10:50:26.6(3)    & 56700 & 0.190067107649(3)   & 0.19546(12)   & 231.1(9)  & 1  &   0.7 &  0.06 &   4.5 \\ [2pt]
J1922+1131  & 19:22:50.27(3)    & +11:31:58.5(6)    & 56700 & 0.56207429041(3)    & 0.0257(10)    & 335(3)    & 1  &   2.4 &  0.13 &  23.0 \\ 
J1925+1720  & 19:25:27.031(4)   & +17:20:27.31(8)   & 56700 & 0.0756589884532(11) & 10.46782(11)  & 223.3(15) & 4  &   0.3 &  0.07 &  2.2 \\ 
J1928+1443  & 19:28:06.79(5)    & +14:43:11.3(12)   & 56700 & 1.01073895346(7)    & 0.210(8)      & 101(5)    & 1  &   3.4 &  0.17 &  41.6 \\ 
J1931+1439  & 19:31:40.433(18)  & +14:39:38.7(4)    & 56700 & 1.77922557495(7)    & 6.328(2)      & 243(16)   & 1  &   1.1 &  0.08 &  64.2 \\ 
J1932+1500  & 19:32:46.307(15)  & +15:00:22.2(4)    & 56700 & 1.86433186749(10)   & 0.459(6)      & 90.5(18)  & 1  &   1.0 &  0.19 &  50.7 \\ [2pt]
J1940+2245  & 19:40:27.644(10)  & +22:45:46.62(20)  & 56700 & 0.258911996567(5)   & 12.71254(19)  & 222.4(13) & 1  &   1.3 &  0.15 &  13.0 \\ 
J1948+2333  & 19:48:19.317(3)   & +23:33:03.23(5)   & 56050 & 0.5283521633015(16) & 13.57843(4)   & 198.2(8)  & 2  &   1.9 &  0.28 &   5.2 \\ 
J1954+2407  & 19:54:00.374(3)   & +24:07:14.00(4)   & 56700 & 0.1934045707829(15) & 1.05616(7)    & 80.5(4)   & 3  &   0.4 &  0.10 &  1.9 \\ 
J2004+2653  & 20:04:59.02(4)    & +26:53:40.4(6)    & 56700 & 0.66587861963(4)    & 0.0567(12)    & 160(4)    & 2  &   2.4 &  0.06 &  14.8 \\ 
\enddata
\tablenotetext{a}{Figures in parentheses are uncertainties in the last digit quoted.}
\tablenotetext{b}{The value of DM has been adjusted for the effects of
interstellar scatter broadening (Section~\ref{sec:profiles}).}
\end{deluxetable}

\begin{deluxetable}{crrccccccc}
\tabletypesize{\scriptsize}
\tablewidth{0pt}
\tablecaption{\label{table:derived}Derived Parameters of 29 Pulsars}
\tablehead{
PSR & \multicolumn{1}{c}{$l$}        & \multicolumn{1}{c}{$b$}        & \multicolumn{1}{c}{DM}              & \multicolumn{1}{c}{$d$\tablenotemark{a}}   & \multicolumn{1}{c}{$L_{1400}$}    && \multicolumn{1}{c}{$\log\,t_c$} &$\log\,B$  & $\log\,\dot{E}$ \\
    & \multicolumn{1}{c}{$(^\circ)$} & \multicolumn{1}{c}{$(^\circ)$} & \multicolumn{1}{c}{(pc\,cm$^{-3}$)} & \multicolumn{1}{c}{(kpc)} & \multicolumn{1}{c}{(mJy kpc$^2$)} && \multicolumn{1}{c}{(log\,yr)}   & (log\,G) & (log\,erg/s)   
}
\startdata
J0611+1436 & 195.38 &  $-$2.00 &   45.7 &   1.5  &     2.4 &&  6.03 &  12.0 &  33.9 \\ 
J1851+0233 &  35.18 &     1.23 &  608.2 &  11.0  &     9.6 &&  6.40 &  11.9 &  33.3 \\ 
J1854+0319 &  36.18 &     0.94 &  480.2 &   8.5  &    12.3 &&  8.26 &  11.3 &  30.9 \\ 
J1855+0306 &  36.17 &     0.48 &  634.3 &   9.4  &     5.4 &&  6.57 &  12.5 &  31.8 \\ 
J1857+0300 &  36.27 &     0.07 &  691.0 &   9.8  &     4.7 &&  6.66 &  12.2 &  32.4 \\ [2pt]
J1858+0319 &  36.71 &  $-$0.09 &  283.7 &   6.1  &     2.2 &&  8.13 &  11.5 &  30.8 \\ 
J1859+0603 &  39.27 &     0.93 &  382.1 &   7.4  &     8.7 &&  6.70 &  12.0 &  32.7 \\ 
J1900+0438 &  38.07 &     0.17 &  630.5 &   9.5  &    10.3 &&  6.19 &  12.0 &  33.6 \\ 
J1901+0234 &  36.37 &  $-$1.05 &  404.3 &   7.4  &     7.3 &&  5.78 &  12.7 &  33.1 \\ 
J1901+0459 &  38.49 &     0.09 & 1112.3 &$\geq$20.0 &$\geq$45.2 &&  5.95 &  12.6 &  33.0 \\ [2pt]
J1901+0511 &  38.71 &     0.08 &  410.1 &   7.1  &     2.3 &&  6.46 &  13.0 &  31.0 \\ 
J1903+0654 &  40.50 &     0.39 &  328.8 &   6.8  &     4.9 &&  6.07 &  12.5 &  32.9 \\ 
J1906+0509 &  39.29 &  $-$1.08 &   99.5 &   3.5  &     0.8 &&  6.08 &  12.2 &  33.5 \\ 
J1907+0255 &  37.34 &  $-$2.19 &  256.9 &   6.1  &     5.4 &&  7.64 &  11.6 &  31.6 \\ 
J1907+0631 &  40.51 &  $-$0.48 &  428.6 &   7.9  &    15.5 &&  4.05 &  13.1 &  35.7 \\ [2pt]
J1907+0859 &  42.71 &     0.66 &  190.0 &   5.2  &     1.8 &&  6.64 &  12.5 &  31.8 \\ 
J1908+0833 &  42.47 &     0.17 &  700.1 &  12.8  &    15.8 &&  6.61 &  12.0 &  32.8 \\ 
J1909+1148 &  45.49 &     1.42 &  199.8 &   5.7  &     2.8 &&  7.99 &  11.3 &  31.5 \\ 
J1913+0657 &  41.64 &  $-$1.71 &  142.2 &   4.6  &     1.1 &&  6.85 &  12.3 &  31.8 \\ 
J1913+1050 &  45.09 &     0.08 &  231.1 &   6.0  &     2.2 &&  7.19 &  11.3 &  33.1 \\ [2pt]
J1922+1131 &  46.76 &  $-$1.60 &  335.1 &   9.1  &    10.5 &&  8.54 &  11.1 &  30.8 \\ 
J1925+1720 &  52.18 &     0.59 &  223.3 &   6.9  &     3.4 &&  5.06 &  12.0 &  36.0 \\ 
J1928+1443 &  50.18 &  $-$1.22 &  100.9 &   4.3  &     3.1 &&  7.88 &  11.7 &  30.9 \\ 
J1931+1439 &  50.54 &  $-$2.00 &  242.9 &   7.7  &     4.7 &&  6.65 &  12.5 &  31.6 \\ 
J1932+1500 &  50.97 &  $-$2.07 &   90.5 &   3.7  &     2.5 &&  7.81 &  12.0 &  30.4 \\ [2pt]
J1940+2245 &  58.63 &     0.13 &  222.4 &   7.4  &     8.3 &&  5.51 &  12.3 &  34.5 \\ 
J1948+2333 &  60.21 &  $-$1.04 &  198.2 &   7.1  &    13.7 &&  5.79 &  12.4 &  33.6 \\ 
J1954+2407 &  61.37 &  $-$1.87 &   80.5 &   3.9  &     1.5 &&  6.46 &  11.7 &  33.8 \\ 
J2004+2653 &  65.03 &  $-$2.53 &  159.7 &   6.3  &     2.3 &&  8.27 &  11.3 &  30.9 \\ 
\enddata
\tablenotetext{a}{Values predicted based on $l$, $b$, and DM, using the NE2001 electron density model of \cite{cl02}.}
\end{deluxetable}

\begin{deluxetable}{lc}
\tabletypesize{\scriptsize}
\tablewidth{0pc}
\tablecaption{\label{table:1932}Observed and derived parameters of the binary pulsar PSR~J1932+1500\tablenotemark{a}}
\tablehead{                                      & PSR~J1932+1500    }
\startdata
Right Ascension (J2000)                          & 19:32:46.307(16)  \\
Declination (J2000)                              & +15:00:22.2(4)     \\
Epoch of pulsar period (MJD)                     & 56700                       \\
Pulsar rotation period (s)                       & 1.8643318675(2)             \\
Pulsar rotation frequency (s$^{-1}$)             & 0.53638518841(3)            \\
First derivative of pulsar frequency (s$^{-2}$)  & $-0.132(2) \times 10^{-15}$ \\ [2pt]
\hline
Binary Period (d)                                & 198.9251(2)    \\
Epoch of Periastron (MJD)                        & 56634.898(14)  \\
Orbital semi-major axis (s)                      & 76.9255(6)     \\
Eccentricity                                     & 0.028928(14)   \\
Longitude of periastron ($^\circ$)               & 292.13(2)      \\ [2pt]
\hline
Binary mass function				 & 0.0123 M$\odot$             \\
Spin-down age (Myr)                              & 64.4                        \\
Spin-down energy ($\mbox{erg/s}$)                & $2.5 \times 10^{30}$        \\
Inferred Magnetic Field (G)                      & $1 \times 10^{12}$          \\
Distance (kpc)\tablenotemark{b}                  & 3.7                         \\
\enddata
\tablenotetext{a}{Figures in parentheses are uncertainties in the last digit quoted.}
\tablenotetext{b}{Value predicted based on $l$, $b$, and DM, using the NE2001 electron density model of \cite{cl02}.}
\end{deluxetable}

\begin{deluxetable}{llrlcl}
\tabletypesize{\scriptsize}
\tablewidth{0pc}
\tablecaption{\label{table:wide-orbits}Wide-orbit binary
pulsars. Together with PSR~J1932+1500 reported here, these are all 
known Galactic binary pulsars taken from the ATNF
Pulsar Catalogue \citep{mhth05} which have orbital period
greater than 50~d and minimum companion mass less than 0.5~M$_\odot$.
The two pulsars with anomalously high values of eccentricity and which
are discussed in the text are highlighted in bold font.}
\tablehead{
\multicolumn{1}{c}{PSR} & \multicolumn{1}{c}{$P$(s)} & \multicolumn{1}{c}{$P_b$(days)} &
\multicolumn{1}{c}{Ecc} & \multicolumn{1}{c}{$M_c$(M$_\odot)$} & \multicolumn{1}{c}{$B$(TG)}}
\startdata
 B0820+02     &   0.864873 &  1232.404 & 0.01187  &  0.19 & 0.304      \\
 J1711$-$4322 &   0.102618 &   922.471 & 0.002376 &  0.20 & 0.0529     \\
 J0407+1607   &   0.025702 &   669.070 & 0.000936 &  0.19 & 0.00144    \\
 J2016+1948   &   0.064940 &   635.024 & 0.001480 &  0.29 & 0.00516    \\
 J0214+5222   &   0.024575 &   512.040 & 0.005328 &  0.41 & 0.00274    \\ [2pt]
 B1800$-$27   &   0.334415 &   406.781 & 0.000507 &  0.14 & 0.0766     \\
 {\bf J1822$-$0848} &   {\bf 2.504518} &   {\bf 286.830} & {\bf 0.05896}  &  {\bf 0.32} & {\bf 1.02}       \\
 {\bf J1932+1500}   &   {\bf 1.864332} &   {\bf 198.925} & {\bf 0.028924} &  {\bf 0.32} & {\bf 0.0529}     \\
 J1640+2224   &   0.003163 &   175.461 & 0.000797 &  0.25 & 0.000096   \\
 J1708$-$3506 &   0.004505 &   149.133 & 0.000244 &  0.16 & 0.000230   \\ [2pt]
 J1643$-$1224 &   0.004622 &   147.017 & 0.000505 &  0.12 & 0.000296   \\
 J2302+4442   &   0.005192 &   125.935 & 0.000503 &  0.29 & 0.000266   \\
 J1529$-$3828 &   0.008486 &   119.675 & 0.000168 &  0.16 & 0.000484   \\
 B1953+29     &   0.006133 &   117.349 & 0.000330 &  0.18 & 0.000432   \\
 J1853+1303   &   0.004092 &   115.654 & 0.000023 &  0.24 & 0.000191   \\ [2pt]
 J1751$-$2857 &   0.003915 &   110.746 & 0.000128 &  0.19 & 0.000212   \\
 J2229+2643   &   0.002978 &    93.016 & 0.000255 &  0.12 & 0.000067   \\
 J1935+1726   &   0.004200 &    90.764 & 0.000175 &  0.22 &    --      \\
 J1850+0124   &   0.003560 &    84.950 & 0.000069 &  0.25 & 0.000199   \\
 J1737$-$0811 &   0.004175 &    79.517 & 0.000053 &  0.07 & 0.000184   \\ [2pt]
 J2019+2425   &   0.003935 &    76.512 & 0.000111 &  0.31 & 0.000168   \\
 J1125$-$5825 &   0.003102 &    76.403 & 0.000257 &  0.26 & 0.000440   \\
 J1455$-$3330 &   0.007987 &    76.175 & 0.000170 &  0.25 & 0.000446   \\
 J1713+0747   &   0.004570 &    67.825 & 0.000074 &  0.28 & 0.000230   \\
 J1910+1256   &   0.004984 &    58.467 & 0.000230 &  0.19 & 0.000222   \\ [2pt]
 J2033+1734   &   0.005949 &    56.308 & 0.000128 &  0.19 & 0.000260   \\
 J0614$-$3329 &   0.003149 &    53.585 & 0.000180 &  0.28 & 0.000238   \\
 J1825$-$0319 &   0.004554 &    52.630 & 0.000193 &  0.18 & 0.000178   \\
 J1844+0115   &   0.004186 &    50.646 & 0.000257 &  0.14 & 0.000214   \\
\enddata
\end{deluxetable}

\begin{deluxetable}{lcc}
\tabletypesize{\scriptsize}
\tablewidth{0pc}
\tablecaption{Parameters of the glitches of PSRs~J0611+1436 and J1907+0631\tablenotemark{a} \label{table:glitches}}
\tablehead{                                      & PSR~J0611+1436                        & PSR~J1907+0631}
\startdata
R.A. (J2000)                                     & 06:11:18.649(13)   & 19:07:03.82(2)      \\
Decl. (J2000)                                    & +14:36:52(4)       & +6:31:18.9(6)       \\
Pulsar rotation frequency (s$^{-1}$)             & 3.6991898114(6)              & 3.0897763136(2)               \\
Pulsar frequency first derivative (s$^{-2}$)     & $-51.37(1) \times 10^{-15}$  & $-4316.60(1) \times 10^{-15}$ \\
Epoch of pulsar frequency (MJD)                  & 55818                        & 56378.03                      \\ [2pt]                  
Glitch epoch (MJD)                               & 55817.8                      & 56985                         \\
Glitch frequency increment (s$^{-1}$)                      & $2.0624(2)\times 10^{-5}$    & $6.495(2) \times 10^{-6}$     \\
Glitch frequency first derivative increment (s$^{-2}$)     & $-0.03(2)\times10^{-15}$     & $-20.9(3) \times 10^{-15}$    \\ [2pt]
\\
Spin-down age (kyr)                              & 1070                         & 11.3                          \\
Spin-down energy ($\mbox{erg/s}/10^{32}$)        & 80                           & 5300                          \\
Inferred Magnetic Field (G$/10^{12}$)            & 1.05                         & 12.2                          \\
\enddata
\tablenotetext{a}{Figures in parentheses are uncertainties in the last digit quoted.}
\end{deluxetable}

\end{document}